\documentstyle[epsfig]{mntt}

\newif\ifAMStwofonts

\ifoldfss
  \ifCUPmtlplainloaded \else
    \NewTextAlphabet{textbfit} {cmbxti10} {}
    \NewTextAlphabet{textbfss} {cmssbx10} {}
    \NewMathAlphabet{mathbfit} {cmbxti10} {} 
    \NewMathAlphabet{mathbfss} {cmssbx10} {} 
  \fi
  \ifAMStwofonts
    \ifCUPmtlplainloaded \else
      \NewSymbolFont{upmath} {eurm10}
      \NewSymbolFont{AMSa} {msam10}
      \NewMathSymbol{\upi}     {0}{upmath}{19}
      \NewMathSymbol{\umu}     {0}{upmath}{16}
      \NewMathSymbol{\upartial}{0}{upmath}{40}
      \NewMathSymbol{\leqslant}{3}{AMSa}{36}
      \NewMathSymbol{\geqslant}{3}{AMSa}{3E}

      \let\leq=\leqslant 
      \let\geq=\geqslant 
    \fi
  \fi
\fi 

\ifnfssone
  \newmathalphabet{\mathit}
  \addtoversion{normal}{\mathit}{cmr}{m}{it}
  \addtoversion{bold}{\mathit}{cmr}{bx}{it}
  \def\textbfit{\protect\txtbfit}
  
  \long\def\txtbfit#1{{\fontfamily{cmr}\fontseries{bx}\fontshape{it}%
    \selectfont #1}}

  \newmathalphabet{\mathbfit} 
  \addtoversion{normal}{\mathbfit}{cmr}{bx}{it}
  \addtoversion{bold}{\mathbfit}{cmr}{bx}{it}
  \newmathalphabet{\mathbfss} 
  \addtoversion{normal}{\mathbfss}{cmss}{bx}{n}
  \addtoversion{bold}{\mathbfss}{cmss}{bx}{n}
  \ifAMStwofonts
    \ifCUPmtlplainloaded \else
      \UseAMStwoboldmath
      \makeatletter
      \new@mathgroup\upmath@group
      \define@mathgroup\mv@normal\upmath@group{eur}{m}{n}
      \define@mathgroup\mv@bold\upmath@group{eur}{b}{n}
      \edef\UPM{\hexnumber\upmath@group}
      \new@mathgroup\amsa@group
      \define@mathgroup\mv@normal\amsa@group{msa}{m}{n}
      \define@mathgroup\mv@bold\amsa@group{msa}{m}{n}
      \edef\AMSa{\hexnumber\amsa@group}
      \makeatother
      \mathchardef\upi="0\UPM19
      \mathchardef\umu="0\UPM16
      \mathchardef\upartial="0\UPM40
      \mathchardef\leqslant="3\AMSa36
      \mathchardef\geqslant="3\AMSa3E

      \let\leq=\leqslant 
      \let\geq=\geqslant 
    \fi
  \fi
\fi 

\ifnfsstwo
  \def\textbfit{\protect\txtbfit}
  
  \long\def\txtbfit#1{{\fontfamily{cmr}\fontseries{bx}\fontshape{it}%
    \selectfont #1}}

  \DeclareMathAlphabet{\mathbfit}{OT1}{cmr}{bx}{it}
  \SetMathAlphabet\mathbfit{bold}{OT1}{cmr}{bx}{it}
  \DeclareMathAlphabet{\mathbfss}{OT1}{cmss}{bx}{n}
  \SetMathAlphabet\mathbfss{bold}{OT1}{cmss}{bx}{n}
  \ifAMStwofonts
    \ifCUPmtlplainloaded \else
      \DeclareSymbolFont{UPM}{U}{eur}{m}{n}
      \SetSymbolFont{UPM}{bold}{U}{eur}{b}{n}
      \DeclareSymbolFont{AMSa}{U}{msa}{m}{n}
      \DeclareMathSymbol{\upi}{0}{UPM}{"19}
      \DeclareMathSymbol{\umu}{0}{UPM}{"16}
      \DeclareMathSymbol{\upartial}{0}{UPM}{"40}
      \DeclareMathSymbol{\leqslant}{3}{AMSa}{"36}
      \DeclareMathSymbol{\geqslant}{3}{AMSa}{"3E}

      \let\leq=\leqslant 
      \let\geq=\geqslant 
    \fi
  \fi
\fi 

\ifCUPmtlplainloaded \else
  \ifAMStwofonts \else 
    \def\upi{\pi}
    \def\umu{\mu}
    \def\upartial{\partial}
  \fi
\fi

\newcommand{\etal}  {et~al.}
\newcommand{\iras}  {{\it IRAS\/}}
\newcommand{\bootes}{Bo\"otes}
\newcommand{\potent}{{\sc potent}}
\newcommand{\wb}    {{\sc wall builder}}
\newcommand{\vf}    {{\sc void finder}}

\newcommand{\hmpc}  {h^{-1}\,\mbox{Mpc}}
\newcommand{\rmax}  {r_{\mathrm{max}}}
\newcommand{\dmax}  {d_{\mathrm{max}}}
\newcommand{\rvol}  {r_o}
\newcommand{\phmgt} {\phantom{$>$}}  
\newcommand{\sect}[1]{Section~\ref{#1}}
\newcommand{\fig}[1]{Fig.~\ref{#1}}

\title[A catalogue of the IRAS voids]
      {A catalogue of the voids in the \textbfit{IRAS\/} 1.2-Jy survey}
\author[H.~El-Ad, T.~Piran and L.~N.~da~Costa]
       {H.~El-Ad,$^1$\thanks{e-mail: {\tt eladh@astro.huji.ac.il}}
        T.~Piran,$^1$\thanks{e-mail: {\tt tsvi@shemesh.fiz.huji.ac.il}} and 
        L.~N.~da~Costa$^{2,3}$\thanks{e-mail: {\tt ldacosta@eso.org}}\\
  $^1$Racah Institute of Physics, The Hebrew University,
      Jerusalem, 91904 Israel\\
  $^2$European Southern Observatory, Karl-Schwarzschild Stra{\ss}e 2,
      D-85748 Garching bei M\"unchen, Germany\\
  $^3$Observat\'orio Nacional, Rua Gen. Jos\'e Cristino 77, 
      S\~ao Cristov\~ao, Rio de Janeiro, Brazil}

\date{Accepted 1996 December 16.
      Received 1996 December 6;
      in original form 1996 August 12}

\pagerange{\pageref{firstpage}--\pageref{lastpage}}

\pubyear{1997}

\begin{document}

\maketitle

\label{firstpage}

\begin{abstract}
  
  Using the \vf\ algorithm we have compiled a catalogue of voids in
  the \iras\ 1.2-Jy sample. The positions of the voids correspond well
  to the underdense regions seen in the \iras\ smoothed density map.
  However, since in our analysis no smoothing is used, all structures
  appear much sharper: walls are not smeared and the voids are not
  artificially reduced by them. Therefore the current method based on
  the point distribution of galaxies is better suited to determine the
  diameter of voids in the galaxy distribution. We have identified 24
  voids, covering more than 30 per cent of the volume considered. By
  comparing the results with equivalent random catalogues we have
  determined that 12 voids are significant at a $0.95$ confidence
  level, having an average diameter of $ 40 \pm 6 \hmpc $. Our results
  serve not only for charting the cosmography of the nearby Universe,
  but also to give support to the results recently obtained with the
  SSRS2 sample, suggesting a void-filled Universe. Moreover, our
  results indicate that the voids detected have a similar scale,
  demonstrating that both optically selected and \iras-selected
  galaxies delineate the same large-scale structures.

\end{abstract}


\section{Introduction}

A remarkable feature of the distribution of galaxies is the existence
of large regions apparently devoid of luminous matter. Since the
discovery of the \bootes\ void \cite{kir81} it was realized that the
existence of large voids in the galaxy distribution could impose
additional constraints to models of large-scale structure (LSS). Since
then complete surveys like the CfA2 \cite{gh89} and SSRS2 \cite{dc94},
which densely sample the nearby galaxy distribution, have shown that
voids are a major feature of the LSS. These surveys show that not only
large voids exist, but more importantly -- that they occur frequently
(at least judging by eye), suggesting a compact network of voids
filling the entire volume. The impact that these findings have in
discriminating models for the origin of LSS has recently been
addressed by Blumenthal \etal\ \shortcite{blu92}, Dubinski \etal\ 
\shortcite{du93}, Piran \etal\ \shortcite{pi93} and van de Weygaert \&
van Kampen \shortcite{wey93}.

As mentioned above, until recently the description of a void-filled
Universe with a characteristic scale of $ 50 \hmpc $ relied solely on
the visual impression of redshift maps. In order to make a more
quantitative analysis we have developed an algorithm for the automatic
detection of voids in three-dimensional surveys. The main features of
the algorithm are as follows.
\begin{enumerate}
\item It is based on the point-distribution of galaxies, not
  introducing any smoothing scale which destroys the sharpness of the
  observed features.
\item It allows for a population of galaxies in voids, recognizing
  that voids need not be completely empty.
\item It tries to avoid the artificial connection between neighbouring
  voids through small breaches in the walls, realizing that walls in
  the galaxy distribution need not be homogeneous as small-scale
  clustering will always be present.
\end{enumerate}

The method has been recently applied to the SSRS2 sample of galaxies
\cite{el96}. Some 12 significant voids with density contrast $ \sim
-0.9 $ were detected with an average diameter of $ 37 \pm 8 \hmpc $
and comprising roughly 40 per cent of the surveyed volume, clearly
supporting earlier qualitative claims. Unfortunately, the advantage of
the dense sampling of the galaxy distribution attained with the SSRS2,
which allows for a larger range of significant voids to be detected,
is offset by the effects of the geometry of the survey.

In order to overcome this limitation we consider here the \iras\ 
1.2-Jy survey \cite{fi95}. Although much sparser than the optical
surveys, the \iras\ sample provides essentially a full sky coverage
minimizing boundary effects. The trade-off is between boundary effects
and the statistical significance of the voids. The \iras\ data also
provide a suitable bench mark as they have been used to derive the
smoothed density field \cite{st95}, and they probe a volume comparable
to that used to determine the density field of the underlying mass
distribution from reconstruction methods based on the measured galaxy
peculiar velocity field (Dekel 1994, Freudling, da~Costa \& Pellegrini
1994, da~Costa \etal\ 1996). Recently, Dekel \etal\ \shortcite{de93}
compared the density field recovered by the \potent\ method with the
smoothed density field of \iras\ galaxies. Both of these fields will
be used below (\sect{discussion}) to compare our results.

Earlier statistical works analysing the \iras\ data (Fisher \etal\ 
1995 and references therein) rarely addressed the voids. An exception
are the works on the Void Probability Function (VPF): Bouchet \etal\ 
\shortcite{bo93} calculated the VPF for various volume-limited \iras\ 
subsamples; Watson \& Rowan-Robinson \shortcite{rr93} incorporated
into the VPF analysis the effects of the selection-function, and
applied it to the one-in-six 0.6-Jy QDOT sample. Both works show how
the VPF departs from Poisson statistics. In addition, several
individual voids were pointed out within the \iras\ sample (cf.\ the
recent review by Strauss \& Willick, chapter~4): among these were the
Local Void \cite{tu87} and the Sculptor Void \cite{dc88}. However, no
objective algorithm has ever been applied to compile a void catalogue
for the whole survey.

We use here an improved version of the \vf\ algorithm (paper~I), to
compile a void catalogue for the \iras\ survey. Our focus here is
two-fold.
\begin{enumerate}
\item Cosmography of the nearby Universe: we chart the individual
  voids, objectively identified within a sphere of radius $ 80 \hmpc
  $, and compare this picture to the one depicted by other techniques.
\item Void statistics: we derive the average diameter and other
  properties of the significant voids, and compare them to those
  derived in paper~I for the SSRS2\@. The possible implications
  regarding galaxy biasing are also discussed.
\end{enumerate}

We review the \vf\ algorithm in \sect{algosect}. The sample we use is
presented in \sect{sampsect}. In \sect{voidsect} we give a pictorial
description of the void distribution, and describe the statistical
properties of the voids that we measure. In \sect{discussion} we
compare several reconstruction methods, and discuss the implications
of our results. Finally, a summary of our main conclusions is
presented in \sect{sumsect}.

\section{The {\sevensize\bf VOID FINDER} algorithm}
\label{algosect}

Since the \vf\ algorithm used for the \iras\ data is a somewhat
improved version of the one used in paper~I for the analysis of the
SSRS2, we include here a description of the updated algorithm. A
complete description can be found in El-Ad \& Piran \shortcite{ep97}.
The algorithm is based on a model in which the main features of the
LSS of the Universe are voids and walls. The walls are thin structures
characterized by a high density of galaxies, separating the voids.
These are underdense regions, but are not completely empty, being
populated by a relatively small number of void galaxies.

\begin{figure*}
  \psfig{file=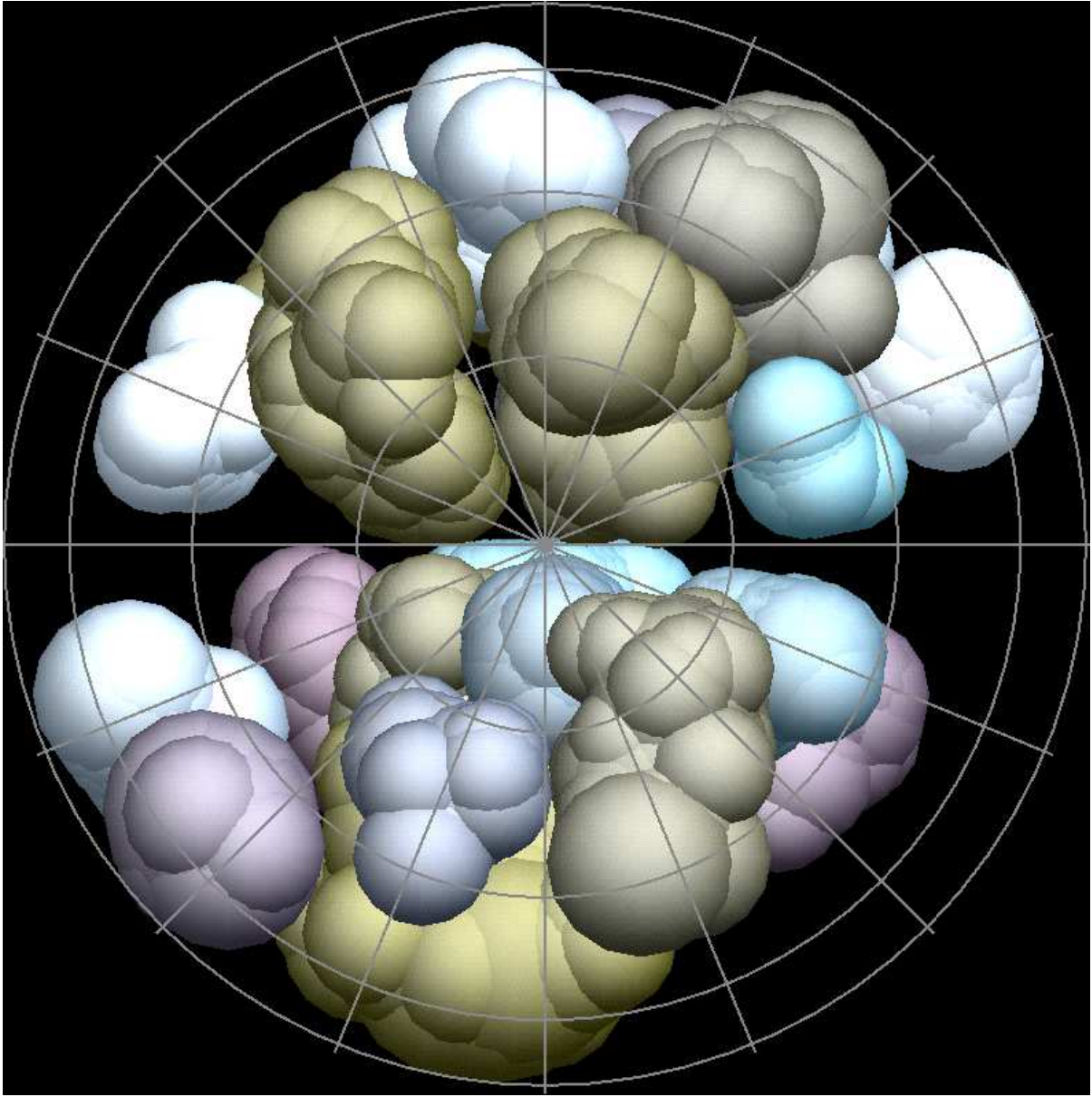, width=.9\linewidth}
\caption[Three-dimensional view of the voids in the \iras\ survey]
{Three-dimensional view of the voids in the \iras\ survey. All 24
  voids are included in this image. The ZOA runs horizontally across
  the image. The area at the left, near the ZOA, with no voids,
  corresponds to the Great Attractor. The absence of voids from the
  lower, right-hand part of the image is due to the Cetus wall and the
  PP supercluster. 3D graphics by Shai Ayal.}
\label{shaifig}
\end{figure*}

The \vf\ algorithm initially locates the voids containing the largest
empty spheres. Following iterations locate the smaller voids, and --
when appropriate -- enlarge the volumes of the older voids. This
identification scheme implies the following definition for a void: it
is a continuous volume that does not contain any (wall) galaxies, and
is thicker than an adjustable limit. Spheres that are devoid of
galaxies are used as building blocks for the voids. A single void is
composed of as many superimposing spheres as required for covering all
of its volume. The algorithm is iterative, with subsequent iterations
searching for voids using a finer {\em void resolution}, which is
defined as the diameter $d_i$ of the minimal sphere used for
encompassing a void during the $i$th iteration. The spheres for
covering a void are picked up in two stages: the {\em identification
  stage}, followed by {\em consecutive enhancements}.
We will now describe these stages in detail.

The {\em identification stage\/} identifies the central parts of the
void. Usually, this is enough for covering about half of the actual
volume, but we focus (at this stage) on identifying a certain void as
a separate entity, rather than trying to capture all of its volume.
The central parts of a void are covered using spheres with diameters
in the range $ \xi \dmax < d \leq \dmax $, with $\dmax$ denoting the
diameter of the largest sphere of a void and $\xi$ being the {\em
  thinness parameter}. The thinness parameter controls the flexibility
allowed while encompassing the central parts of a void. Setting $ \xi
= 1 $ would leave us with only the largest sphere in the void, while
lowering $\xi$ allows the addition of more spheres. If the void is
composed of more than one sphere (as is usually the case), then each
sphere must intersect at least one other sphere with a circle wider
than the minimal diameter $ \xi \dmax $. We have taken $ \xi = 0.85 $,
which allows for enough flexibility -- still without accepting
counter-intuitive void shapes. A lower $\xi$ reduces the total number
of the voids, with a slow increase in their total volume. Once a group
of such intersecting spheres has been dubbed a void, it will not be
merged with any other group.

After the central part of a void is identified, we {\em consecutively
  enhance\/} its volume, in order to cover as much of the void volume
as possible using the current void resolution. These additional
spheres need not adhere to the $\xi$ thinness limitation: during each
subsequent iteration, we scan the immediate surroundings of each of
the voids already identified. If we find additional empty spheres that
intersect with the void, then these are added to the void. We scan for
enhancing spheres of a certain diameter only {\em after\/} scanning
for new voids with that diameter. In this way we do not falsely break
apart individual voids, and we do not prevent the identification of
truly new voids.

Since the average galaxy number density decreases with depth, as only
the brighter galaxies are observable at greater distances, we must
apply corrections to the algorithm in order to minimize these effects.
The correction used by the \vf\ is to scale the diameters of the
spheres by the selection function, thus accepting only relatively
larger spheres in the sparser regions of the survey.

The full algorithm includes also a pre-selection stage -- the \wb\ --
that identifies wall galaxies (which are used later to define the
voids) and field galaxies (which are ignored). However, since the
\iras\ sample is relatively sparse, we have taken a more conservative
approach, and considered all the galaxies while identifying the voids.
Hence, the \iras\ voids presented below are {\em completely empty}.
Comparison of these results with voids found after filtering field
galaxies indicates that the filtering process affects the cosmography
(some voids are merged), but it has only a small effect on the void
statistics (see \sect{voidsect}).

We define a {\em wall galaxy\/} as a galaxy that has at least $n$
other wall galaxies within a sphere of radius $\ell_n$ around it. The
radius $\ell_n$ is derived based on the statistics of the distance to
the $n$th nearest neighbour. In our analysis we use $ n = 3 $. The
\wb\ corrects for the selection function in a similar way to the \vf:
we determine $\ell_3$ in the volume-limited region of the sample (see
below), beyond which we scale it by the selection function.

\begin{figure*}
  \psfig{file=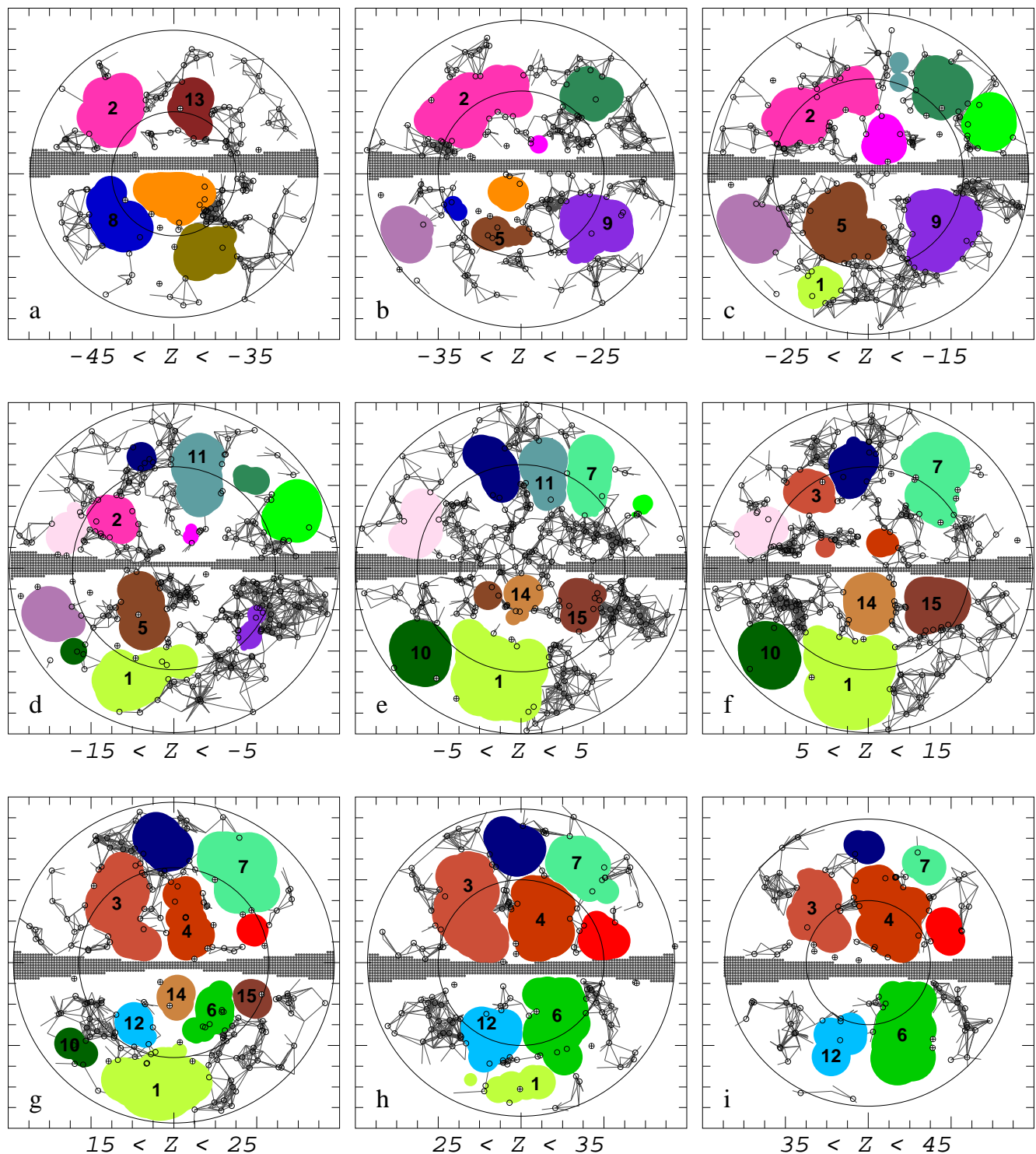, width=\textwidth}
\caption[\iras\ voids and walls in slices parallel to the SG plane]
{\iras\ voids and walls in slices parallel to the SG plane. The shaded
  areas mark the intersection of the centre of the given plane with
  the three-dimensional voids. Darker shading implies a more
  significant void. Only the fifteen $ p > 0.8 $ voids are numbered
  (see Table~\ref{voidtable}). The excluded ZOA is indicated along $ Y
  = 0 $. The depicted galaxies extend $ 5 \hmpc $ above and below the
  plane. Wall galaxies are marked as by `$\circ$', and field galaxies
  by `$\oplus$'. Note that {\em all\/} the galaxies are located
  outside the voids -- galaxies that seem to be in a void appear so
  due to the two-dimensional projection. As a visual aid for
  highlighting the walls, we have drawn lines connecting between all
  pairs of wall galaxies closer than the properly scaled wall
  separation distance $\ell_3$. These artificial connections are not
  used in our analysis. If the SGZ plane drawn is perpendicular to a
  wall, we see a thin dense feature -- e.g., as seen in panel (h)\@.
  If the projection plane coincides with a plane of a wall, we see a
  wide dense feature. This is evident in the SG plane (panel e), where
  the Cetus wall and PP occupy a large fraction of the slice. The
  inner circle at $ \rvol = 50 \hmpc $ marks the volume-limited region
  of our sample. The outer circle marks the boundary of the sample, at
  $ \rmax = 80 \hmpc $.}
\label{slicefig}
\end{figure*}

\section{The sample}
\label{sampsect}

The \iras\ survey contains 5321 galaxies complete to a flux limit of
1.2-Jy \cite{fi95}. We applied corrections for the computed peculiar
velocities, to obtain the real-space distribution of the galaxies. In
the void analysis, we have limited ourselves to galaxies extending out
to $ \rmax = 80 \hmpc $, and created a semi--volume-limited sample
consisting of galaxies brighter than \mbox{$ l_{60} \geq 3.59 \times
  10^{30} h^{-2} {\mathrm{erg\,s}}^{-1} {\mathrm{Hz}}^{-1} $} at $ 60
\umu \mathrm{m} $, corresponding to a depth $ \rvol = 50 \hmpc $. The
selection function $\phi$ drops to 22 per cent at $\rmax$. The final
sample consists of 1876 galaxies. The 1531 faint galaxies that were
eliminated in order to create the volume-limited region of the survey
are not used when processing the walls nor the voids. However, after
the survey is analysed and the voids located, we examine the locations
of these faint galaxies (see \sect{statsect}).

The sky coverage of the \iras\ is almost complete ($87.6$ per cent),
with the galactic plane region $|b| < 5 \degr $ constituting most of
the excluded zones. Various schemes (e.g., that of Yahil \etal\ 1991)
have been used to extrapolate the density field to the Galactic plane,
but these are not directly applicable to our analysis. Thus, when
looking for voids we avoid the zone of avoidance (ZOA), treating it as
a rigid boundary practically cutting the \iras\ sample into two
halves. Since the ZOA cuts across voids this scheme divides some voids
to two and eliminates others. However, it is the most conservative
method, and therefore the results for the volumes of the voids should
be considered as lower limits. We estimate the effect of this method
by examining the opposite approach in which the ZOA is treated as if
it is a part of the survey, applying no corrections. The ZOA is
nowhere wider than the minimal void resolution used, so it does not
create new voids by itself. Therefore the effect of including the ZOA
is to overestimate the size of voids near it, because it allows the
merging of a couple of voids and the expansion of other voids into the
region. Still the overall effect on the void statistics is limited
(see \sect{statsect}).

Areas lacking sky coverage in the Point Source Catalog constitute most
of the remaining excluded zones. These were processed as if they were
included in the \iras, as their effect is rather negligible. However,
when the voids we find include these regions, that is indicated (see
\sect{cosmosect}).

The \wb\ analysis of the \iras\ galaxy distribution located 95 per
cent of the galaxies within walls. Each wall galaxy was required to
have at least three other wall galaxies within a sphere of radius $
\ell_3 = 10.2 \hmpc $ around it.
We find that the walls occupy at most $ \sim 25 $ per cent of the
examined volume. This corresponds to an average wall overdensity of at
least \mbox{$ \delta \rho / \rho \approx 4 $}. Note that here we have
used a sample somewhat deeper than $\rmax$ so galaxies located near
the boundary of our sample are not mistakenly recognized as field
galaxies.

\section{The \textbfit{IRAS\/} voids}
\label{voidsect}

\subsection{Cosmography}
\label{cosmosect}

Applying our most conservative approach to analyse the \iras\ data --
i.e., including the field galaxies and avoiding the ZOA -- we have
identified 24 voids of which 12 are statistically significant at a
$0.95$ confidence level (see \sect{statsect}). \fig{shaifig} depicts a
three-dimensional view of the \iras\ voids. \fig{slicefig} shows the
voids and the walls, in nine planes parallel to the supergalactic (SG)
plane at $ 10 \hmpc $ intervals. In general, some of the voids shown
are smaller than their actual size -- because of the way we treat the
ZOA, or because the field galaxies were not removed from the analysis.
Both effects imply that our estimates of the size of voids are likely
to be lower limits.

In the SG plane (\fig{slicefig}, panel e) one recognizes void~10 as
the Sculptor Void \cite{dc88}, located below the
Pavo--Indus--Telescopium part ($ Y < 0 $) of the Great Attractor (GA),
seen here to be composed of several substructures. Adjacent to it we
find void~1, stretching parallel to the Cetus wall. These two voids
are separated only by a few field galaxies. If we filter them out, the
two would merge to form one huge void, equivalent in volume to a $ d =
62 \hmpc $ sphere occupying most of that part of the skies. Voids 1
and 10 are limited by the $\rmax$ boundary of our sample, so they
could prove to be larger still.

The area above the Perseus--Pisces (PP) supercluster (up to the Great
Wall near Coma, at $ Y = 70 \hmpc $) is occupied by two voids: 7 and
11\@. If the field galaxies are filtered first, these two voids merge.
Also note in this area the minor void ($ p = 0.21 $) located below the
Coma supercluster, at $( X = -7, Y = 54 )$: this void (extending to
the $ Z > 0 $ panels) corresponds to the largest void found in the CfA
survey \cite{lgh86}.

The closest void we found (void~14) can be seen in the centre of this
panel, just below the Local Supercluster. Another clear, and rather
nearby, void in the SG plane is void~15, in front of PP\@. A minor
void can be viewed beyond the $ Y > 0 $ section of the GA, at $( X =
-51, Y = 19 )$.

Above the SG plane (\fig{slicefig}, panels f--i) we see the extensions
of the SG plane voids, as well as some additional voids -- voids 3, 4,
6 and 12\@. Void~4 is the Local Void \cite{tu87}. Below the SG plane
(\fig{slicefig}, panels a--d) we note void~2 (just above the
Hydra--Centaurus supercluster), and voids 5 and 9. Voids 8 and 13
extend around $ Z = -40 $ and can be seen in panel (a)\@. We should
point out that voids 1, 2, 5, 6, 7, 11 and 14 include areas lacking
sky coverage in the PSC\@.

\begin{figure}
  \psfig{bbllx=65, bblly=295, bburx=535, bbury=731, clip=,
    file=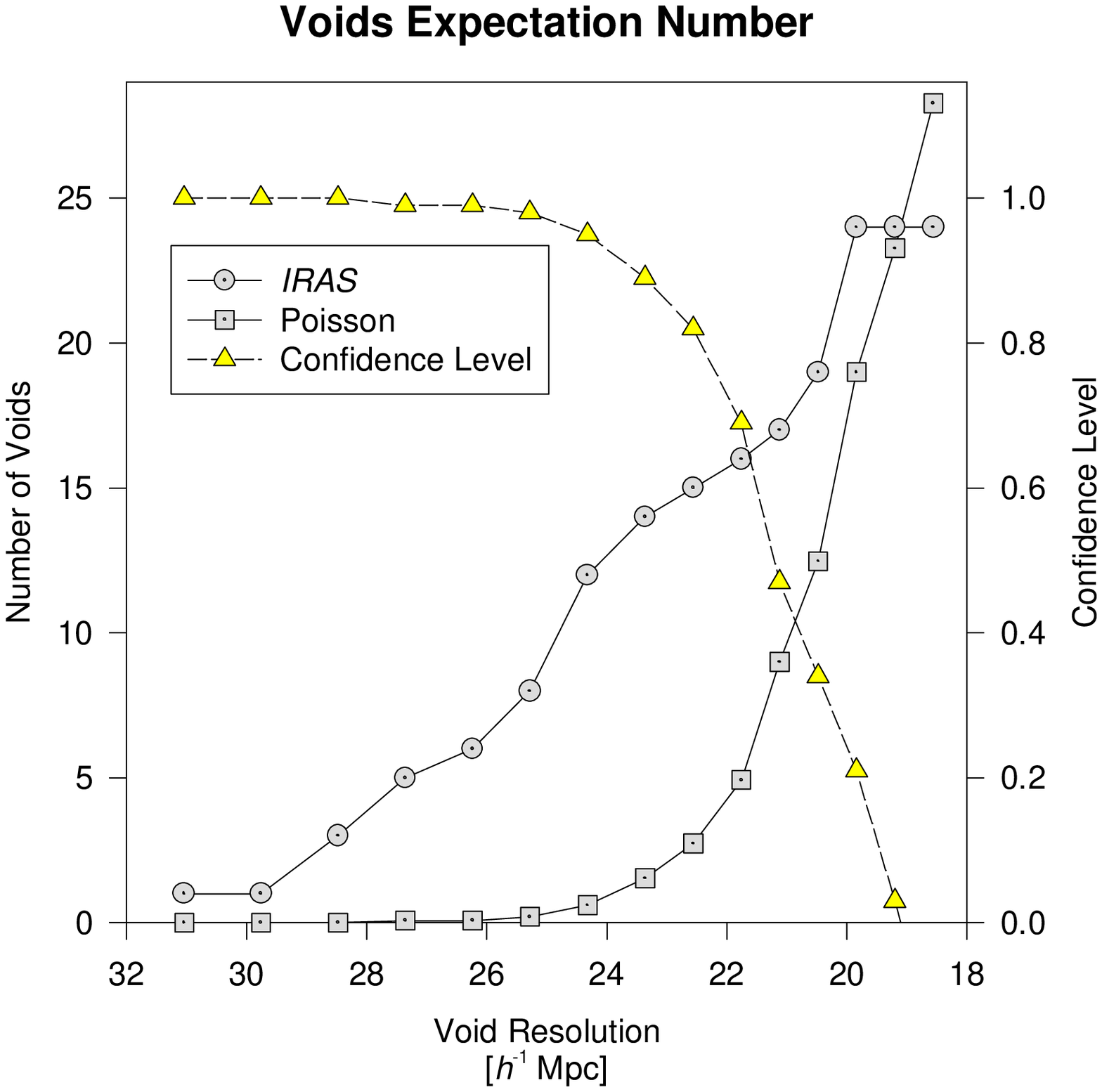, width=\linewidth}
\caption[The accumulated number of voids as a function of the void resolution]
{The accumulated number of voids as a function of the void resolution,
  for the \iras\ sample and for random catalogues. The derived
  confidence level $p$ is also indicated.}
\label{accfig}
\end{figure}

\begin{table*}
\begin{minipage}{154mm}
\caption{Locations and properties of the voids in the \iras\ survey.}
\label{voidtable}
\begin{tabular}{rccccrrrcc}
\hline
 & Statistical & Equivalent & Total & \multicolumn{4}{c}{Location of Centre} & Largest & \\
 & Confidence & Diameter & Volume & \multicolumn{4}{c}{(Supergalactic coordinates)} & Sphere's & Identification \\
\cline{5-8}
 & Level & {[$\hmpc$]} & {[$h^{-3}\,\mbox{KMpc}^3$]} & \multicolumn{1}{c}{$r$} & \multicolumn{1}{c}{$X$} & \multicolumn{1}{c}{$Y$} & \multicolumn{1}{c}{$Z$} & Fraction & \\
 & (1) & (2) & (3) & (4) & \multicolumn{1}{c}{(5)} & \multicolumn{1}{c}{(6)} & \multicolumn{1}{c}{(7)} & (8) & \multicolumn{1}{c}{(9)} \\
\hline
 1 & $>$0.99 & 51.0 & 69.9 & 55.2 & -10.7 & -53.8 &   6.1 & 0.34 & \\ 
 2 & $>$0.99 & 43.8 & 44.4 & 49.6 & -25.3 &  31.4 & -28.9 & 0.33 & \\ 
 3 & $>$0.99 & 44.5 & 45.9 & 46.0 & -24.8 &  26.7 &  28.1 & 0.28 & \\ 
 4 & \phmgt 0.99 & 45.0 & 47.4 & 46.5 &   8.7 &  24.7 &  38.4 & 0.25 & Local Void \\ 
 5 & \phmgt 0.99 & 36.0 & 24.4 & 32.0 & -13.0 & -23.9 & -16.9 & 0.46 & \\ 
 6 & \phmgt 0.99 & 41.4 & 37.3 & 51.5 &  17.0 & -32.2 &  36.4 & 0.30 & \\ 
 7 & \phmgt 0.98 & 43.5 & 43.3 & 57.1 &  31.2 &  44.9 &  16.5 & 0.31 & \\ 
 8 & \phmgt 0.98 & 39.5 & 32.6 & 60.4 & -25.8 & -22.7 & -49.7 & 0.50 & \\ 
 9 & \phmgt 0.95 & 36.0 & 24.4 & 49.8 &  35.9 & -25.6 & -23.0 & 0.35 & \\ 
10 & \phmgt 0.95 & 33.6 & 19.9 & 63.3 & -48.0 & -40.9 &   6.0 & 0.81 & Sculptor Void \\ 
11 & \phmgt 0.95 & 32.0 & 17.2 & 48.6 &  11.8 &  46.6 &  -6.9 & 0.52 & \\ 
\vspace{6pt}
12 & \phmgt 0.95 & 31.5 & 16.5 & 49.9 & -15.6 & -35.7 &  31.3 & 0.46 & \\ 
13 & \phmgt 0.89 & 40.3 & 34.5 & 62.8 &  14.2 &  29.3 & -53.7 & 0.47 & \\ 
14 & \phmgt 0.89 & 28.8 & 12.7 & 19.0 &   0.7 & -16.4 &   9.6 & 0.58 & \\ 
15 & \phmgt 0.82 & 30.4 & 14.6 & 37.6 &  32.4 & -17.0 &   8.6 & 0.42 & Perseus--Pisces Void \\ 
\hline
\end{tabular}
\end{minipage}
\end{table*}

\begin{figure*}
\noindent
\begin{minipage}{.469\linewidth}
  \raggedleft {\Huge a} \psfig{bbllx=108, bblly=540, bburx=234,
    bbury=666, clip=, file=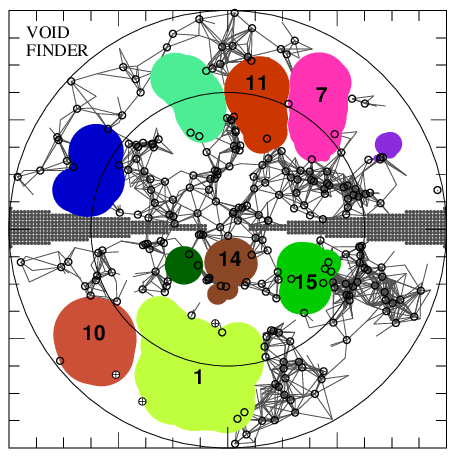, height=220pt,
    width=220pt}
\end{minipage}\hspace{1pt}%
\noindent%
\begin{minipage}{.469\linewidth}
  \raggedright \psfig{bbllx=80pt, bblly=205pt, bburx=564pt,
    bbury=690pt, clip=, file=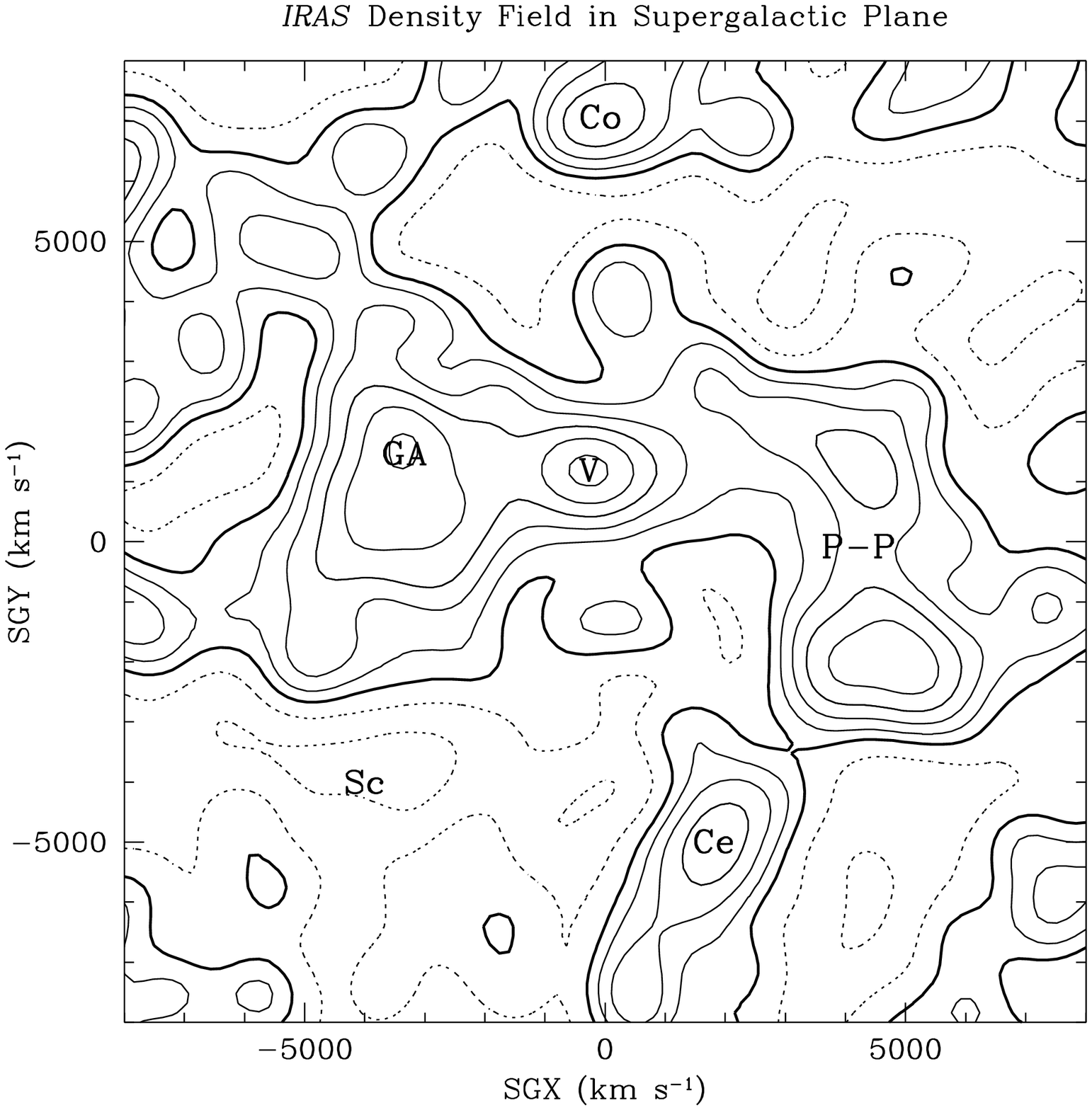, height=220pt,
    width=220pt} {\Huge b}
\end{minipage}
\noindent
\begin{minipage}{.469\linewidth}
  \raggedleft {\Huge c} \psfig{bbllx=98pt, bblly=223pt, bburx=565pt,
    bbury=690pt, clip=, file=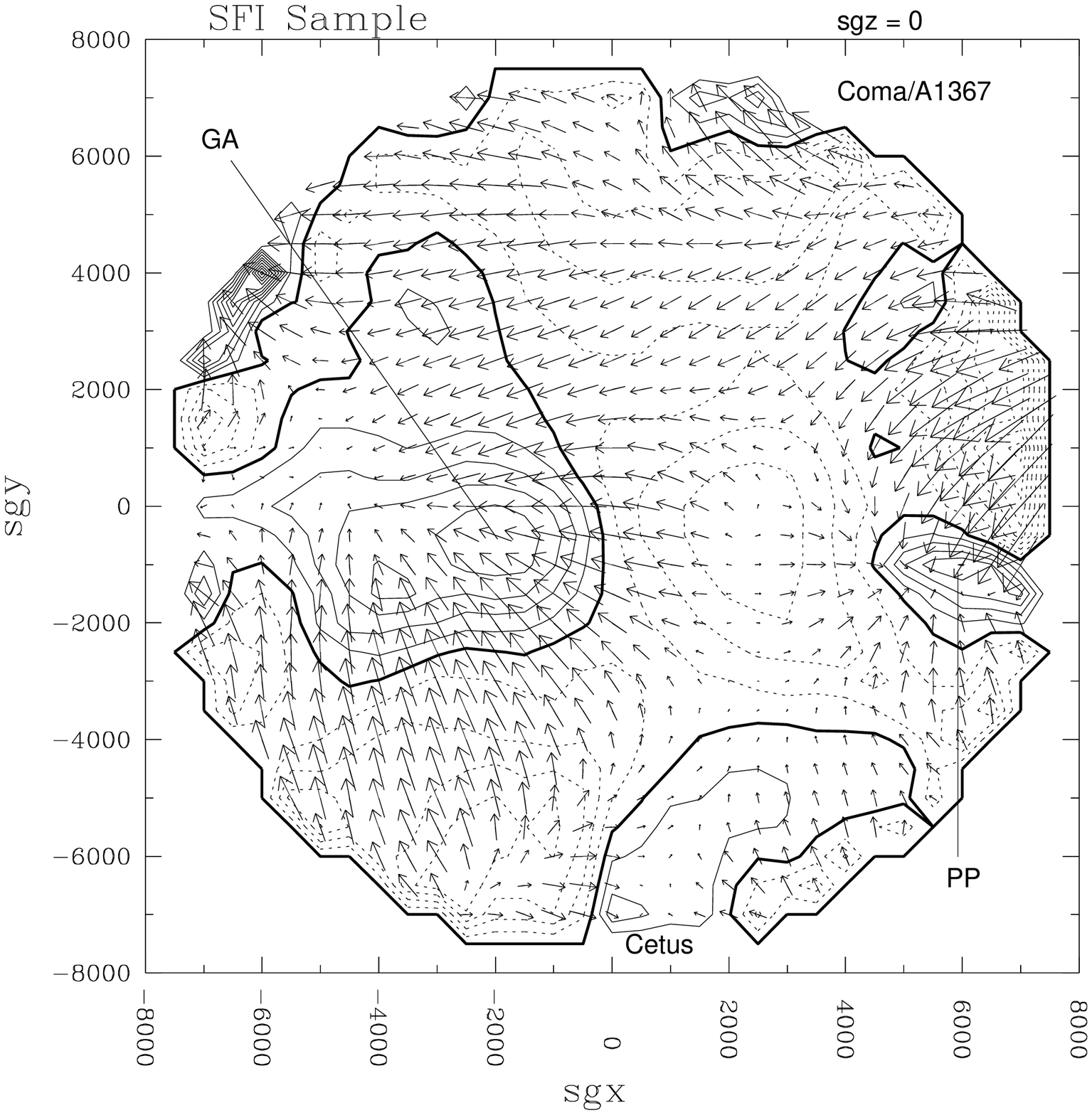, height=220pt,
    width=220pt}
\end{minipage}\hspace{1pt}%
\noindent%
\begin{minipage}{.469\linewidth}
  \raggedright \psfig{bbllx=84pt, bblly=174pt, bburx=529pt,
    bbury=620pt, clip=, file=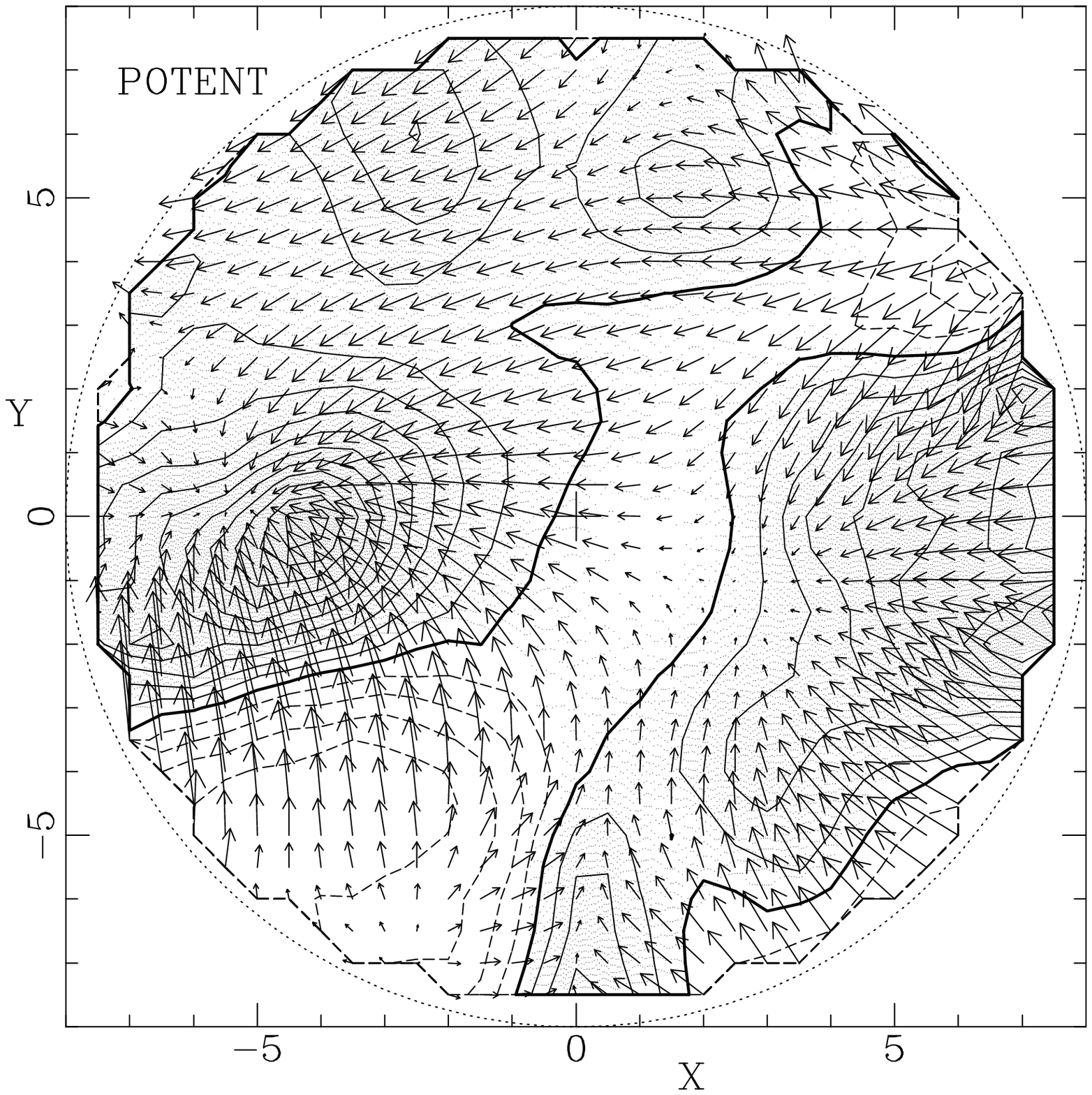, height=220pt,
    width=220pt} {\Huge d}
\end{minipage}
\caption[The supergalactic plane as depicted by various techniques]
{The supergalactic plane extending out to $ 80 \hmpc $, as depicted by
  various techniques. Panel (a): the locations of the voids and the
  walls, using the \vf\ algorithm. See the caption of \fig{slicefig}
  for details. Panel (b) (Strauss \& Willick 1995): the real-space
  smoothed density field of \iras\ galaxies, using $ 5 \hmpc $
  Gaussian smoothing, extrapolating into the ZOA\@. The density field
  is obtained by a self-consistent correction for peculiar velocities
  with $\beta = 1$. Reproduced by permission of Michael Strauss.
  Panel (c) (da Costa \etal\ 1996): the reconstructed velocity and
  density fields obtained from the SFI sample, using $ 9 \hmpc $
  Gaussian smoothing. The arrows give the $X$--$Y$ components of the
  three-dimensional velocity field. The contours are of $\delta$,
  spaced at $0.2$ intervals. The heavy solid line indicates $\delta =
  0$. Panel (d) (Dekel 1994; Dekel \etal, in preparation): the
  smoothed velocity field and the resultant density field as recovered
  by \potent\ from the Mark~III data, using $ 12 \hmpc $ Gaussian
  smoothing. Reproduced by permission of Avishai Dekel.}
\label{sgpfig}
\end{figure*}

\begin{figure*}
  \psfig{file=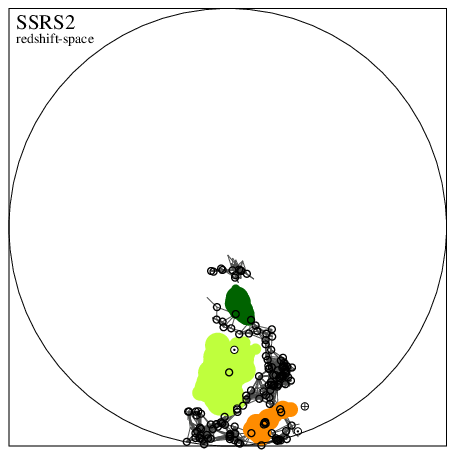, width=0.4\linewidth} \hspace{20pt}
  \psfig{file=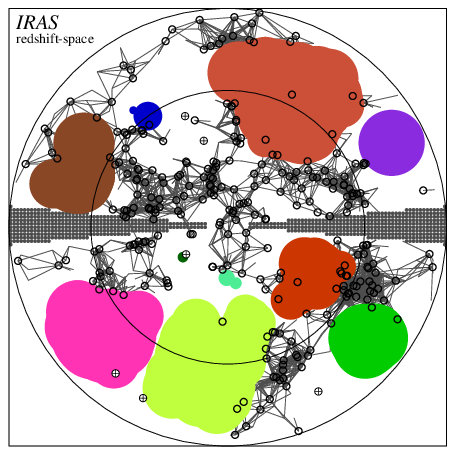, width=0.4\linewidth}
\caption[\iras\ and SSRS2 voids in the SG plane]
{Voids in the SG plane: here we compare the redshift-space voids of
  the SSRS2 (left) and the \iras\ (right). The denser sampling of the
  SSRS2 is evident. Similar voids are found in the overlapping regions
  of the surveys. For the \iras\ sample compare also the real- and
  redshift-space distribution: the \iras\ voids in redshift-space are
  larger, and the dense structures appear much more collapsed than in
  real-space (\fig{sgpfig}, panel a).}
\label{iras-ssrs2-fig}
\end{figure*}

\subsection{Void statistics}
\label{statsect}

In order to assess the statistical significance of the voids, we have
created random distributions built to mimic the geometry and density
of the true sample. Averaging over the random catalogues, we derive
the expected number of voids in Poisson distributions as a function of
the void resolution (see \fig{accfig}). We will denote by $
N_{\mathrm{Poisson}}(d) $ the number of voids in a Poisson
distribution that contain a sphere whose diameter is $d$. The same
quantity for the actual distribution will be denoted $
N_{\mathit{IRAS}}(d) $\@. Note that $d$ scales with the selection
function, correcting for the reduced density as $r$ increases. Using
these void counts, we define the {\em confidence level\/} as \[ p(d) =
1 - \frac{N_{\mathrm{Poisson}}(d)}{N_{\mathit{IRAS}}(d)} \mathrm{.}\]
The closer $p$ is to unity, the less likely such a void could appear
in a random distribution. We consider voids with $p > 0.95$ as
statistically significant. In the \iras\ sample, we have identified 12
such voids, and only these are considered in the calculations below.
Our statistical confidence level should not be interpreted as the
usual $2\sigma$ or $3\sigma$ grade, as the statistics we use has a
different interpretation. For example, having 12 voids at a $0.95$
confidence level means that {\em on average\/} one finds $0.6$ such
voids in the random catalogues. So, perhaps one of these 12 voids
could be attributed to a random process, but we do not know which one.

Three additional voids have a moderate confidence level $ 0.8 < p <
0.95 $. The fact that void~15 ($ p = 0.82 $) is a well recognized void
(the PP void) hints that all voids satisfying $ p > 0.8 $ are worth
mentioning. It is the sparseness of the \iras\ 1.2-Jy that prevents us
from establishing higher confidence levels: with denser surveys we
expect that the physical reality of the lower confidence level voids,
as well as their formal statistical confidence level, will both be
established. Our experience regarding the transition to denser surveys
has taught us that the new galaxies rarely appear in the voids, where
most of them concentrate in the already dense areas (e.g., the
extension of the CfA survey to $ m_B \leq 15.5 $ -- see de~Lapparent
\etal\ 1986). Thus an \iras\ extension to 0.6-Jy will probably not
show smaller voids, and will enable higher statistical confidence
levels. At the void resolution of $ 18.6 \hmpc $, $
N_{\mathrm{Poisson}}(d) $ exceeds $ N_{\mathit{IRAS}}(d) $, and we
terminate the void search as $p$ now vanishes. At this stage, 24 voids
were found in the \iras.

In Table~\ref{voidtable} we list the locations and properties of the
fifteen $p > 0.8$ voids. The 12 most significant voids ($p > 0.95$)
are listed in the upper part of the table. Column (1) lists the
statistical confidence level $p$ of each void. The diameters given in
column (2) are of a sphere with the same volume as the whole void, as
is listed in column (3). The centre of the void is defined as its
centre-of-(no)-mass; the distance to it, and its exact location, are
given in columns (4)--(7). In column (8) we indicate the fraction of
the total volume of the void covered by the single largest sphere it
contains. Finally, column (9) identifies some of the voids.

The average size of the 12 significant voids in the \iras\ sample as
estimated from the equivalent diameters is \mbox{$ \bar{d} = 40 \pm 6
  \hmpc $}, consistent both with the higher $ 50 \hmpc $ eye estimates
of Geller \& Huchra \shortcite{gh89} and da Costa \etal\ 
\shortcite{dc94}, and with the lower $ 38 \hmpc $ estimate obtained
from the first zero-crossing of the SSRS2 correlation function
\cite{gl95} and from our void analysis of the SSRS2 (paper~I). The
increase in average void diameter in the \iras\ compared to the SSRS2
($\sim 5$ per cent) is due to the relatively narrow angular limits of
the latter survey.

The 12 most significant \iras\ voids occupy 22 per cent of the
examined volume; considering all 24 voids, the volume is 32 per cent.
If we consider only the volume-limited region of our sample, where
there are no distortions caused by the $\rmax$ boundary of the survey
(only by the ZOA), the void volume reaches 46 per cent. We have also
examined the void distribution in redshift-space. As expected, voids
in redshift-space are typically bigger than their real-space
counterparts. The total void volume in redshift-space is $\sim 20$ per
cent larger that that in real-space, and the average diameter of the
significant voids in redshift-space is $ 44 \hmpc $ (compare
\fig{sgpfig}, panel a, with \fig{iras-ssrs2-fig}, right panel).

After the voids were located we examined the locations of the
previously eliminated faint galaxies. Only 204 (13 per cent) of the
faint galaxies are located within the voids, in agreement with the
identification of the voids based on the brighter galaxies. However,
as found in the SSRS2 (paper~I), there is a notable increase in the
number of faint galaxies in the voids compared with the number of
brighter galaxies.

What is the effect of the limitations we have imposed in our void
analysis? As stated above, the treatment of the ZOA as a rigid
boundary and the consideration of only empty voids cause us to
interpret the results derived in this way as a lower limit. An upper
limit is derived by taking the opposite approach, this time including
the ZOA and filtering the field galaxies. Each factor alone
corresponds to an increase in the average void diameter of 5 to 15 per
cent. Together the effect is $\sim 20$ per cent, yielding an upper
limit for this sample of $ \bar{d} = 48 \hmpc $. A similar increase
occurs in the total void volume. When filtering the field galaxies the
voids are not empty, now having an average underdensity of \mbox{$
  \delta \rho / \rho \approx -0.9 $}, as found for the SSRS2 in
paper~I\@.

\section{Discussion}
\label{discussion}

\fig{sgpfig} depicts the SG plane, as analysed using four different
methods: the \vf\ technique, the \iras\ density field \cite{st95} and
the reconstructed velocity and density fields from the SFI sample
\cite{dc96} and from the Mark~III catalogue (Dekel 1994; Dekel \etal,
in preparation). The voids and walls identified by our algorithm
indeed correspond to the under- and overdense regions in the \iras\ 
density field respectively. Comparison with the SFI sample as well
indicates a good agreement for most of the voids. On the other hand,
the comparison with the Mark~III map reveals several conflicts, where
for instance voids 7, 11 and 15 are replaced by overdense features in
the Mark~III reconstruction.

Most of the overdense regions, walls and filaments are narrower than $
10 \hmpc $. The smoothing scale used for creating the density fields
spreads the originally thin structures over wider regions, extending
into the underdense volumes. This has the effect of giving a false
impression of a rather blurred galaxy distribution, where prominent
overdense structures are separated by small underdense regions. The
true picture is very different: there is a sharp contrast between the
thin overdense structures which occupy only the lesser part of the
volume, and the large voids. The notion of a void-filled universe
cannot be avoided in this picture.

Comparison of panels (a) and (c) of \fig{sgpfig} also demonstrates
that the voids delineated by galaxies correspond remarkably well with
the underdense regions in the reconstructed mass density field derived
from peculiar velocity measurements. This supports the idea that the
observed voids in redshift surveys represent true voids in the mass
distribution.

An additional comparison was performed directly between two \vf\ 
reconstructions: we examined the void distribution in the region where
the SSRS2 sample (paper~I) overlaps the \iras\ sample.
\fig{iras-ssrs2-fig} depicts the redshift-space voids in the SG plane,
for the \iras\ sample (right panel) and for the corresponding part of
the SSRS2 sample (left panel). In this region we find three of the 12
significant voids identified in the SSRS2 sample. The corresponding
\iras\ voids are $\sim 10$ per cent larger (in diameter) than the
SSRS2 ones, since they are not bounded by narrow angular limits as are
the SSRS2 voids. The increase indicated here is larger than that
indicated earlier (\sect{statsect}), where the whole surveys were
compared. When only individual voids are compared, the voids in the
SSRS2 located at greater distances cannot be taken into account, and
these are less affected by the angular limits of the survey.

The two surveys agree not only regarding the locations of individual
voids in the limited volume where the surveys overlap, but also when
we compare the gross statistical properties of the voids. Both surveys
show a similar void scale of $ \sim 40 \hmpc $, with an average
underdensity \mbox{$\delta \rho / \rho \approx -0.9$}. Although
smaller voids could not appear in our \iras\ analysis (they lack
statistical significance) and larger voids could hardly fit in the
volume we analyse ($ \rmax = 80 \hmpc $), we suggest that this void
scale is indeed a characteristic physical scale.
\begin{enumerate}
\item Our analysis method reproduces all known voids in the regions
  examined, and it agrees with both the \iras\ density field and the
  \potent\ reconstruction based on the SFI sample.
\item A similar void scale appears in both the \iras\ and the SSRS2
  samples, withstanding the inherent differences between the two
  surveys, regarding sky coverage, galaxy selection (optical/{\it
    IR\/}) and density. Further still, the \iras\ galaxies represent a
  special galaxy class, possibly biased relative to the optical
  galaxies \cite{la88}. The two surveys agree not only statistically,
  but also on an individual void basis. Although the SSRS2 is much
  denser than the \iras, the voids in it are not smaller.
\item An eye examination of the largest survey available today, the
  Las-Campanas Redshift Survey \cite{sl96} indicates again a similar
  void scale -- the voids do not seem to be much larger although the
  survey is.
\end{enumerate}
The characteristic void scale found in the two surveys supports the
idea mentioned earlier, that the voids are also devoid of dark matter
and have formed gravitationally \cite{pi93}.

\section{Summary}
\label{sumsect}

We have used the \vf\ algorithm to derive a catalogue of voids in the
\iras\ redshift survey. Due to the relatively sparse sampling of this
survey, we have taken a conservative approach in our analysis, looking
for completely empty voids and avoiding the ZOA\@. As such, the
average void size derived $ \bar{d} = 40 \pm 6 \hmpc $ should be
considered a lower limit for the actual size of the voids.
Nevertheless, thanks to the nearly full sky coverage of the \iras\ 
sample, it is probably the best-suited redshift survey currently
available for deriving a void spectrum and for charting the nearby
void cosmography. The \vf\ analysis clearly shows the prominence of
the voids in the LSS, not hindered by smoothing of the overdense
regions.

The main features of the LSS of the Universe found in the SSRS2 sample
(paper~I), are repeated in the \iras\ sample. Namely, these are as
follows.
\begin{enumerate}
\item Large voids occupying $\sim 50$ per cent of the volume.
\item Walls occupying less than $\sim 25$ per cent of the volume.
\item A void scale of at least $ 40 \hmpc $, with an average
  underdensity of $-0.9$.
\item Faint galaxies do not `fill the voids', but they do populate
  them more than bright galaxies.
\end{enumerate}
This consistency between \iras\ and optically selected galaxies is
based on an objective measurement of the most prominent feature of the
LSS of the Universe, the voids, suggesting that galaxies of different
types delineate equally well the observed voids. Therefore galaxy
biasing is an unlikely mechanism for explaining the observed voids in
redshift surveys. Comparison with the recovered mass distribution
further suggests that the observed voids in the galaxy distribution
correspond well to underdense regions in the mass distribution. If
true this will confirm the gravitational origin of the voids.

\section*{Acknowledgments}
We thank S.~Ayal for helpful discussions and comments, and for
creating the three-dimensional void image. We are grateful to Avishai
Dekel and to Michael Strauss for providing two of the figures.
\fig{sgpfig}(b) was reprinted from Phys.\ Rep., 261, Strauss \&
Willick, The density and peculiar velocity fields of nearby galaxies,
271, 1995 with kind permission of Elsevier Science -- NL, Sara
Burgerhartstraat 25, 1055 KV Amsterdam, The Netherlands.
\fig{sgpfig}(c) has been reproduced, with permission, from
Astrophysical Journal, published by the University of Chicago Press
(\copyright\ 1996 by the American Astronomical Society. All rights
reserved).

\bspp 

\label{lastpage}

\end{document}